\documentclass[12pt]{article}
\usepackage[dvips]{graphicx}
\usepackage{amssymb}
\newcommand{\eq}{\begin{equation}}
\newcommand{\en}{\end{equation}}
\newcommand{\eqa}{\begin{eqnarray}}
\newcommand{\ena}{\end{eqnarray}}

\begin{document}

\begin{titlepage}
\vskip0.5cm
\begin{flushright}
\end{flushright}
\vskip0.5cm
\begin{center}
{\Large\bf A Monte Carlo study of the three-dimensional XY universality class:
Universal amplitude ratios
}
\end{center}
\vskip 0.3cm

\centerline{\sl  Institut f\"ur Theoretische Physik, Universit\"at Leipzig
}
\centerline{\sl  Postfach 100 920, D-04009 Leipzig, Germany}
\centerline{\sl
e--mail: \hskip 0.2cm
 Martin.Hasenbusch@itp.uni-leipzig.de}
\vskip0.4 cm

\begin{abstract}
We simulate lattice models in the three-dimensional XY universality class 
in the low and the 
high temperature phase. This allows us to compute a number of universal 
amplitude ratios with unprecedented precision: $R_{\Upsilon}=0.411(2)$, 
$R_B=2.83(1)$, $R_{\xi}^+=0.3562(10)$ and $R_{\xi}^-=0.850(5)$.
These results can be compared with those obtained  from
other theoretical methods, such as field theoretic methods or the high  
temperature series expansion and also with experimental
results for the $\lambda$-transition of $^4$He. In addition to the XY model,
we study the three-dimensional two-component $\phi^4$ model on the simple cubic 
lattice.  The parameter of the $\phi^4$ model is chosen such that 
leading corrections to scaling are small.
\end{abstract}


\end{titlepage}

\section{Introduction}
In the neighbourhood of a second order phase transition various 
quantities diverge, following power laws.
E.g. in a magnetic system, the magnetic susceptibility
behaves as
\begin{equation}
\label{chipower}
 \chi \simeq C_{\pm}  |t|^{-\gamma} \; ,
\end{equation}
where $\gamma$ is the critical exponent of the magnetic susceptibility,
$C_{+},C_{-}$ are the amplitudes in the high and low temperature phase,
respectively, and $t=(T-T_c)/T_c$ is the reduced temperature.
Critical exponents like $\gamma$ assume 
universal values; i.e. they assume exactly the same value for all systems 
within a given universality class. Following Wilson (see e.g.\cite{WiKo}),
such a universality class is characterized by the dimension of the system, 
the range of the interaction and the symmetry of the order parameter.
In addition to the critical exponents, amplitude ratios like $C_{+}/C_{-}$
are universal, while the value of $C_{+}$ or $C_{-}$ depends on the 
microscopic details of the model. For a review on amplitude ratios see ref.
\cite{PrHoAh91}.

In the present work we compute, using data obtained from Monte 
Carlo simulations of lattice models, the numerical values 
of four  such universal amplitude ratios for the universality class
of the three dimensional XY model. The $\lambda$-transition of 
$^4$He is supposed to share this universality class. 
At temperatures below the transition, $^4$He becomes superfluid.
The  $\lambda$-transition owes its name to the fact that the specific 
heat plotted
as a function of temperature resembles the Greek letter $\lambda$.
The order parameter of the $\lambda$-transition in $^4$He
is the phase of a wave function. Therefore it should share the
XY universality class, which is characterized by the O(2), or equivalently U(1),
symmetry of the order parameter.
The study of the $\lambda$-transition provides by far the most precise experimental
results for universal quantities like critical exponents and amplitude
ratios. Thus this transition gives us a unique opportunity to test the
ideas of the renormalization group and to benchmark  theoretical
methods.
For a review and an outlook on future experiments in space-stations
\footnote{The condition of micro-gravity avoids 
the broadening of the transition due to the gravitational field and hence
allows to access reduced temperatures down to $5 \times 10^{-10}$
\cite{LSNCI-96,lipa2003}.}
see ref. \cite{BaHaLiDu07}.

The present work is the completion of ref. \cite{MHHe}, where we had  
computed the ratio $A_{+}/A_{-}$ of the amplitudes of the specific heat.
Our results can be compared with those obtained by using other theoretical
methods and, which may be even more important, experimental results obtained 
for the $\lambda$-transition of $^4$He. 

The outline of our paper is the following: First we define the models and 
the observables that are measured.  We briefly discuss the Monte Carlo 
algorithm that has been used for the simulation. 
Since the continuous $O(2)$-symmetry  is 
spontaneously broken in the low temperature phase, there is a Goldstone boson. 
As a consequence, the thermodynamic limit is approached with corrections that 
decay as inverse powers of the linear lattice size.
Therefore, extracting estimates for the thermodynamic limit from Monte Carlo data
requires special care. To this end, we summarize the relevant results of chiral
perturbation theory as discussed in refs. \cite{HaLe1990,DiHaNaNi1991,ToYo94}.
Then we present our numerical estimates of various observables in the low and high
temperature phases. Next we compute the amplitude ratios from these observables.
Finally we compare our results with those of other theoretical methods and
experiments.

\section{The models}
We study the $\phi^4$ model on a simple cubic lattice 
with periodic boundary conditions in each of the directions.  
The classical Hamiltonian is given by
\begin{equation}
 H_{\phi^4} = - \beta \sum_{<x,y>} \vec{\phi}_x \cdot \vec{\phi}_y
   + \sum_{x} \left[\vec{\phi}_x^2 + \lambda (\vec{\phi}_x^2 -1)^2   \right]
\;\;,
\end{equation}
where the field variable $\vec{\phi}_x$ is a vector with two real 
components. $<x,y>$ denotes a pair of nearest neighbour sites. 
The sites of the lattice are labelled by $x=(x_1,x_2,x_3)$ with 
$x_i \in \{ 1,2,...,L_i\}$. Throughout we consider lattices with $L_1=L_2=L_3=L$.
Note
that in our convention, following ref. \cite{newXY}, the inverse temperature
$\beta$ is absorbed into the Hamiltonian but does not multiply its second
term. 
The partition function is given by
\begin{equation}
 Z_{\phi^4} = \prod_x \left[\int \mbox{d} \phi_x^{(1)} \int \mbox{d} \phi_x^{(2)} \right]
 \;\; \exp( -H_{\phi^4} ) \;\;.
\end{equation}
The reduced free energy density is then given by $f = - \frac{1}{L^3} \log(Z)$. 

In the limit $\lambda \rightarrow \infty$ the classical XY (plane rotator)
model is recovered.
It is defined by the classical Hamiltonian
\begin{equation}
 H_{XY} = - \beta \sum_{<x,y>} \vec{s}_x \vec{s}_{y} \;\;,
\end{equation}
where $\vec{s}_x$ is a unit-vector with two real components.   
For $\lambda=0$ one gets the exactly solvable Gaussian model.
For $0 < \lambda \le \infty$ the $\phi^4$ model undergoes a second 
order phase transition in the XY universality class; 
see e.g. ref. \cite{BiJa05}.

Power laws like eq.~(\ref{chipower}) are subject to corrections; 
see e.g. ref. \cite{review};
\begin{equation}
 \chi = C_{\pm}  |t|^{-\gamma}  \; 
\left(1 + b_{\pm} t^{\theta} + ...\right) 
\end{equation}
with $\theta \approx 0.5$ for the three-dimensional XY universality class.
Such corrections complicate the determination of universal quantities from Monte Carlo
simulations or high temperature series expansions of lattice models. Correction
to scaling amplitudes like $b_{\pm}$ depend on the parameter $\lambda$ of the
$\phi^4$ model. It is a rather old idea \cite{improved_idea} to search for 
the value of such a parameter, where the leading correction to scaling amplitude
vanishes. Note that the renormalization group predicts that the zero of the leading 
correction amplitude is the same for all quantities.
It has been demonstrated numerically \cite{HaTo99,ourXY} 
that such an improved value $\lambda^*$ indeed exists.
The most accurate determination of the improved value is $\lambda^*=2.15(5)$  
\cite{newXY}.  Previous estimates are 
$\lambda^* = 2.07(5)$ in ref. \cite{ourXY}
and $\lambda^* = 2.10(6)$ in ref. \cite{HaTo99}.

We performed simulations  at $\lambda=2.1$ and $\lambda=2.2$,
where leading corrections to scaling are small.
Following ref. \cite{newXY}  leading corrections to scaling at $\lambda=2.1$
and $\lambda=2.2$ should be at least by a factor of 20 smaller than
in the XY model. Comparing results obtained with $\lambda=2.1$ and
$\lambda=2.2$ allows us to estimate the effect of these small corrections.
In ref. \cite{newXY} the estimates $\beta_c=0.5091503(3)[3]$ and 
$\beta_c=0.5083355(3)[4]$ for the inverse critical temperatures at
$\lambda=2.1$ and $\lambda=2.2$, respectively, were obtained. 
The number quoted in $()$ gives the statistical error, 
while the number given in $[]$ gives the systematic one. 
In the following analysis of our data we will just add up these two numbers.
In order to clearly see effects due to leading corrections to scaling, we 
also have simulated the XY model. 
Recent estimates for the inverse of the critical temperature
are $\beta_c=0.454165(4)$, $0.454167(4)$, $0.4541659(10)$ 
and $0.4541652(11)$ in refs. 
\cite{spain,Bielefeld2,Bloete05,newXY},
respectively. In the analysis of our numerical data we shall assume 
$\beta_c=0.0.4541655(10)$, which is roughly the average of the estimate of 
refs. \cite{Bloete05} and \cite{newXY}.

\subsection{Simulation algorithm}
We have simulated the XY model using the single cluster
algorithm \cite{Wolff}. The $\phi^4$ model was simulated  using
a hybrid of the single cluster algorithm and a local Metropolis algorithm.
The cluster algorithm allows only to change the angle of the field
$\vec{\phi}$. In order to to change the modulus $|\vec{\phi}|$,
Metropolis updates are performed.
Such a hybrid approach was originally proposed by
Tamayo and Brower \cite{BrTa} for the one component $\phi^4$ model.
The generalization to $N$ components is discussed in ref. \cite{HaTo99}.
For details of our implementation, in particular of the local updates 
see ref. \cite{newXY}.

\section{Observables}
In the following we give the precise definitions of the quantities that 
we have measured in our Monte Carlo simulations. To this end we use the 
notation of the $\phi^4$ model. The definitions for the XY model can be obtained 
by replacing $\vec{\phi}$ by $\vec{s}$ in these definitions. 

\subsection{The energy and the specific heat}
In order to study universal quantities it is not crucial how the transition 
line in the $\beta$-$\lambda$ plane is crossed, as long as this path is 
smooth and not tangent to the transition line.  
Here, following computational convenience, we vary $\beta$ at fixed $\lambda$.
Correspondingly we define the energy density as the derivative of the reduced 
free energy with respect to $\beta$. Furthermore we multiply by $-1$ to get 
positive numbers: 
\begin{equation}
 E =  \frac{1}{L^3} 
 \left \langle  \sum_{<x,y>} \vec{\phi}_x \cdot \vec{\phi}_y \right \rangle \;.
\end{equation}
We then define the specific heat as the derivative of the energy density with 
respect to $\beta$: 
\begin{equation}
 C_h = \frac{1}{L^3} 
 \left (\left \langle \left [\sum_{<x,y>} \vec{\phi}_x \cdot \vec{\phi}_y
 \right ]^2 \right \rangle - 
 \left \langle \sum_{<x,y>} \vec{\phi}_x \cdot \vec{\phi}_y \right \rangle^2 \right)  \;.
\end{equation}

\subsection{The magnetisation, the magnetic susceptibility and 
the correlation length}
The magnetisation of a configuration is given by
\begin{equation}
\vec{m} =  \frac{1}{V} \sum_x \; \vec{\phi}_x  \;.
\end{equation}
The second moment correlation length is defined by
\begin{equation}
\xi_{2nd} \;=\; \sqrt{\frac{\chi/F-1}{4 \; \sin(\pi/L)^2}} \;\;\;,
\end{equation}
where
the magnetic susceptibility in the high temperature phase is given by
\begin{equation}
\chi \;=\; \frac{1}{2 V} 
\; \left\langle \left(\sum_x \; \vec{\phi}_x \right)^2 \right\rangle \;,
\end{equation}
where we assume that the magnetisation vanishes in the high temperature phase.
The Fourier transform of the correlation function at the lowest non-zero momentum is
\begin{equation}
F_k \;= \; \frac{1}{2 V} \;   \left \langle
\left |\sum_x \exp\left(i \frac{2 \pi x_k}{L} \right) \vec{\phi}_x \right |^2 
\right \rangle \;\;.
\end{equation}
In order to reduce the statistical error we have averaged $F$  over all
three directions $k=1,2,3$.

\subsection{The helicity modulus $\Upsilon$}
 The helicity modulus $\Upsilon$ gives the reaction of the system under
a torsion. To define the helicity modulus
we introduce rotated boundary conditions in one direction:
For $x_1=L_1$ and $y_1=1$ we replace the term $\vec{\phi}_x  \vec{\phi}_y$ 
in the Hamiltonian by
\begin{equation}
\vec{\phi}_x \cdot R_{\alpha} \vec{\phi}_y =
\phi_x^{(1)} \left(\cos(\alpha) \phi_y^{(1)} + \sin(\alpha) \phi_y^{(2)} \right)
+\phi_x^{(2)}\left(-\sin(\alpha) \phi_y^{(1)} + \cos(\alpha) \phi_y^{(2)}\right)
\end{equation}

The helicity modulus is then given by
\begin{equation}
\label{helidef}
\left . \Upsilon = - \frac{L_1}{L_2 L_3}
\frac{\partial^2 \log Z(\alpha)}{\partial \alpha^2} \right|_{\alpha=0} \;\;.
\end{equation}
Note that we have skipped for convenience a factor $1/(k_B T)$ compared with 
the standard definition. 
The helicity modulus can be directly evaluated in the Monte Carlo simulation.
Following eq.~(3) of ref. \cite{teli89} one gets
\begin{equation}
\label{helimeasure}
\Upsilon = 
\frac{\beta}{L_1 L_2 L_3}
\left \langle \sum_x \vec{\phi}_x \vec{\phi}_{x+(1,0,0)} \right \rangle - 
\frac{\beta^2}{L_1 L_2 L_3}
\left \langle  \left[\sum_x (\phi_x^{(1)} \phi_{x+(1,0,0)}^{(2)} 
    - \phi_x^{(2)} \phi_{x+(1,0,0)}^{(1)}) \right ]^2  \right \rangle
\end{equation}
To arrive at this expression, the torsion by $\alpha$  
is spread over the lattice ; i.e. using a torsion by $\alpha/L_1$ 
at any $x_1$. In order to reduce
the statistical error we have measured the helicity modulus also 
in $x_2$ and $x_3$-direction.

The helicity modulus is of particular interest, since it plays a central 
role in the effective description of the behaviour in the low temperature
phase, as we shall see below. 
Furthermore it can be accessed experimentally in the superfluid phase
of $^4$He. In the literature, the inverse of the helicity modulus, as 
defined here, is called transverse correlation length. 
It is given by  \cite{FiBaJa73}
\begin{equation}
\label{xip}
\frac{1}{\Upsilon} = \xi_{\perp} = \frac{m_4^2 k_b T}{\hbar^2 \rho_{s}} \;,
\end{equation}
where $\rho_{s}$ is the superfluid density and $m_4$ the mass of a $^4$He atom.
The superfluid density can be obtained from measurements of the second 
sound and the specific heat.

\section{The Goldstone mode and finite size effects}
\label{goldstone}
Before we discuss the numerical results in the low temperature phase, 
let us summarize the relevant results from chiral perturbation theory.
These results are derived in ref. \cite{HaLe1990,DiHaNaNi1991} 
for $O(N)$-invariant 
nonlinear $\sigma$ models in three dimensions. Let us briefly recall the 
results for our case $N=2$, which is simpler than the general case. 
Furthermore, we consider only the case of a vanishing external field.

The basic assumption of chiral perturbation theory is that in the broken phase, on scales much larger than the correlation length,
only fluctuations perpendicular to the overall
magnetisation remain as degrees of freedom. Their fluctuations can be described 
by an effective Hamiltonian $H_{eff}[\psi]$, where the field $\psi$ gives the 
angle of the fluctuation. Due to the $O(2)$ invariance 
of the underlying microscopic model, the effective model has to be invariant
under global shifts of the field. I.e. it can only depend on derivatives 
of the field $\psi$. Also by symmetry, it can only depend on even powers 
of odd derivatives.

In the leading, Gaussian 
approximation, the  effective Hamiltonian, using lattice notation, is given by
\begin{equation}
\label{gausseff}
H_{eff}[\psi] = \frac{\beta_{eff}}{2} \sum_{<xy>} (\psi_x - \psi_y)^2 \;.
\end{equation}
In the limit $\beta \rightarrow \infty$, the effective Hamiltonian provides
an approximation of the XY model in a microscopic sense with 
$\beta_{eff}=\beta$.  However, in the neighbourhood of the phase transition, 
it serves (only) as an effective model. This means in particular that the 
relation between $\beta_{eff}$ and $\beta$ is a priori not known.
Hence $\beta_{eff}$ has to be determined from observables that characterize 
the behaviour of the system at large scales.
To this end, the helicity modulus is particularly useful. Plugging in 
the effective Hamiltonian $H_{eff}[\psi]$ into definition~(\ref{helidef})
one obtains $\beta_{eff}=\Upsilon$. Corrections to this relation are due to
higher order terms in the effective Hamiltonian. One gets
\begin{equation}
\label{Upsilonansatz}
\Upsilon(L) = \Upsilon(\infty)\; +\; c_{\Upsilon} \; L^{-3} \;\;,
\end{equation}
for the $L$-dependence of the helicity modulus.
A similar relation holds for the energy density:
\begin{equation}
\label{eneransatz}
 E(L) = E(\infty)\; +\; c_E \; L^{-3} \;\;.
\end{equation}

In the case of the two-point function, a non-trivial effect arises from 
the fact that the spin is a non-linear function of the angle. This has to be taken 
into account in the case of the expectation value of the square magnetisation.
From eq. (2.18) of ref. \cite{ToYo94} we read
\begin{equation}
\label{fssmag}
\langle m^2 \rangle = \Sigma^2 \left[\frac{1}{2} \rho_1^2  + 2 \rho_2 \alpha^2 \right]
\end{equation}
where 
\begin{equation}
\rho_1 = 1 + \frac{1}{2} \beta_1 \alpha + \frac{1}{8} (\beta_1^2 -2 \beta_2)
          \alpha^2  \;\;,\;\;\;\;\; \rho_2 = \frac{1}{4} \beta_2
\end{equation}
with $\beta_1=0.225785$, $\beta_2=0.010608$
and 
\begin{equation}
\alpha= \frac{1}{\Upsilon L} \;\;.
\end{equation}
These results are derived from more general expressions given in section 
10 of ref. \cite{HaLe1990}.  Corrections to eq.~(\ref{fssmag}) are proportional 
to $\alpha^3$. 

\section{Simulations in the low temperature phase}
\label{lowtemp}
We have simulated the $\phi^4$ model at $\lambda=2.1$ and $\lambda=2.2$ and the 
XY model for various values of $\beta$ in the low temperature phase. Most of these simulations
were already carried out in the context of ref. \cite{MHHe}, where only the data
for the energy density were analysed.
For each value of $\beta$ and each model we have simulated a number of lattice sizes
to extrapolate, following the predictions of chiral perturbation theory, 
to the thermodynamic limit. In the case of the $\phi^4$ model, for our smallest
values of $\beta$, we have simulated lattices up to the linear size $L=288$.
Typically we performed $10^5$ to $2 \times 10^5$ measurements for each
value of $\lambda$, $\beta$ and $L$. For each of these
measurements one Metropolis sweep and a few single cluster
updates were performed. In total the simulations
in the low temperature phase took about 1.5 years of CPU time on a 2 GHz
Opteron CPU.

As a typical example, we give in table \ref{phi4b515} the results for 
$\lambda=2.1$ and $\beta=0.515$. 
\begin{table}
\caption{\sl \label{phi4b515}
Results for various lattice sizes $L$ for the 2-component $\phi^4$ model at
$\lambda=2.1$ and $\beta=0.515$.  We give numerical estimates 
for the helicity modulus $\Upsilon$, the energy density $E$ 
and the square of the
magnetisation  $<m^2>$.  $\Sigma^2$ is defined by eq.~(\ref{fssmag}).
For a detailed discussion see the text.
}
\begin{center}
\begin{tabular}{|l|l|l|l|l|}
\hline
  \multicolumn{1}{|c}{$L$}
& \multicolumn{1}{|c}{$\Upsilon$}
& \multicolumn{1}{|c}{$E$}
& \multicolumn{1}{|c}{$\langle m^2 \rangle$} 
& \multicolumn{1}{|c|}{$\Sigma^2/2$} \\
\hline
\phantom{0}32 & 0.05097(14) & 1.01763(8) & 0.09038(8) & 0.07822(7)[2] \\
\phantom{0}48 & 0.04989(14) & 1.01655(5) & 0.08504(6) & 0.07725(5)[2] \\
\phantom{0}64 & 0.04968(14) & 1.01620(3) & 0.08249(5) & 0.07677(5)[1] \\
\phantom{0}96 & 0.04938(15) & 1.01606(2) & 0.08036(6) & 0.07661(6)[1] \\
          128 & 0.04938(14) & 1.01603(1) & 0.07938(4) & 0.07657(4)[1] \\
\hline
\end{tabular}
\end{center}
\end{table}
First we have fitted the helicity modulus $\Upsilon$  and the energy 
density $E$ with the ans\"atze~(\ref{Upsilonansatz},\ref{eneransatz}),
respectively.
Using the data for all lattice sizes given in table \ref{phi4b515}, we 
get $\Upsilon(\infty)=0.049394(8)$ and $E(\infty)=1.016005(8)$. These
values are close to those obtained for the largest lattice size $L=128$
that we have simulated for $\lambda=2.1$ and $\beta=0.515$.
Therefore we regard the extrapolation as save.
In table \ref{lowphi4} we summarize our results for $\Upsilon(\infty)$
and $E(\infty)$ for all values of  $\beta$ that we have  
simulated in the low temperature phase of the $\phi^4$ model at $\lambda=2.1$.
Analogous results for $\lambda=2.2$ and the XY model 
can be found in tables \ref{lowphi4l2.2} and \ref{lowXY}, respectively. 
In addition to the statistical error, we give in parenthesis the differences
$\Upsilon(\infty)-\Upsilon(L_{max})$ and
$E(\infty)-E(L_{max})$ as a rough estimate of the systematic error.
$L_{max}$ is the largest lattice size that is available at a given 
value of $\beta$.
In order to compute the thermodynamic limit of the magnetisation we 
used the results of chiral perturbation theory summarized in 
section \ref{goldstone}. As input, we have taken our result for the 
thermodynamic limit of the helicity modulus. In the last column of 
table  \ref{phi4b515} we give our results for $\Sigma/2$ 
for each value of $L$ using eq.~(\ref{fssmag}). In order to get an estimate
for the thermodynamic limit, we have fitted these results with the 
ansatz $\Sigma(L) = \Sigma(\infty)  + c_S L^3$. Results 
obtained this way, for all values of $\beta$ that have been simulated, 
are given in the last column of table \ref{lowphi4},  \ref{lowphi4l2.2} and 
\ref{lowXY}
for $\lambda=2.1$, $\lambda=2.2$ and the XY model, respectively.

\begin{table}
\caption{\sl \label{lowphi4}
Results for the low temperature phase of the 2-component $\phi^4$ model at 
$\lambda=2.1$. $\beta$ is the inverse temperature, $\Upsilon$ the 
helicity modulus, $E$ is the energy and $\Sigma^2$ the square of the 
magnetisation. For the definition of these quantities 
and a detailed discussion see the text.
}
\begin{center}
\begin{tabular}{|l|l|l|l|}
\hline
  \multicolumn{1}{|c}{$\beta$}
& \multicolumn{1}{|c}{$\Upsilon$}
& \multicolumn{1}{|c}{$E$}
& \multicolumn{1}{|c|}{$\Sigma^2$} \\
\hline
  0.51  & 0.01330(15)[+4] &0.931674(14)[-16] & 0.04020(5)[-7] \\
  0.5105& 0.01818(18)[-31]&0.941232(16)[-13] & 0.05545(5)[-5] \\
  0.511 & 0.02265(17)[-25]&0.950382(11)[-9]  & 0.06905(5)[-4] \\
  0.512 & 0.03036(10)[-17]&0.967852(11)[-24] & 0.09322(5)[-6] \\
  0.513 & 0.03717(11)[-29]&0.984474(10)[-29] & 0.11480(7)[0]  \\
  0.515 & 0.04939(8)[+1] & 1.016005(8)[-27]& 0.15305(7)[-12]\\
  0.52  & 0.07456(7)[-13]& 1.088210(8)[-17]& 0.23338(8)[+4] \\
  0.525 & 0.09628(7)[-3] & 1.15416(1)[-2]  & 0.30099(9)[-14]\\
  0.53  & 0.11565(7)[-15]& 1.21557(2)[-7]  & 0.36133(11)[-16]\\
  0.535 & 0.13349(6)[+6] & 1.27342(2)[-5]  & 0.41614(11)[+2]\\
  0.54  & 0.15035(6)[-7] & 1.32830(2)[-4]  & 0.46694(14)[+14]\\
  0.55  & 0.18145(6)[-8] & 1.43072(3)[-7]  & 0.55824(17)[-33]\\
  0.58  & 0.26340(5)[-18]& 1.69533(6)[-35] & 0.78325(22)[-52]\\
\hline
\end{tabular}
\end{center}
\end{table}

\begin{table}
\caption{\sl \label{lowphi4l2.2}
Same quantities as in table  \ref{lowphi4} for $\lambda=2.2$.
}
\begin{center}
\begin{tabular}{|l|l|l|l|}
\hline
  \multicolumn{1}{|c}{$\beta$}
& \multicolumn{1}{|c}{$\Upsilon$}
& \multicolumn{1}{|c}{$E$}
& \multicolumn{1}{|c|}{$\Sigma^2$} \\
\hline
   0.5095 & 0.01670(16)[+2]  & 0.937817(19)[-18] & 0.04986(5)[-6] \\
   0.51   & 0.02131(17)[0]   & 0.947000(19)[+5]  & 0.06386(5)[-6] \\
   0.511  & 0.02877(16)[-20] & 0.964370(17)[-14] & 0.08840(4)[-3]\\
   0.512  & 0.03602(24)[+17] & 0.980920(19)[-14] & 0.11022(5)[-4] \\
\hline
\end{tabular}
\end{center}
\end{table}

\begin{table}
\caption{\sl \label{lowXY}
Same quantities as in table  \ref{lowphi4} for the standard XY model.
}
\begin{center}
\begin{tabular}{|l|l|l|l|}
\hline
  \multicolumn{1}{|c}{$\beta$}
& \multicolumn{1}{|c}{$\Upsilon$}
& \multicolumn{1}{|c}{$E$}
& \multicolumn{1}{|c|}{$\Sigma^2$} \\
\hline
  0.456 &   0.02043(16)[-11]& 1.01820(1)[-2]& 0.06674(6)[-8] \\
  0.458 &   0.03331(16)[-7] & 1.04568(2)[-3]& 0.11036(6)[-9] \\
  0.46  &   0.04399(9)[-22] & 1.07095(1)[-6] & 0.14648(4)[-16] \\
  0.462 &   0.05334(13)[-18]& 1.09473(2)[-15]& 0.17851(12)[-31]\\
  0.465 &   0.06618(12)[+4] & 1.12832(2)[-9] & 0.22097(11)[-18]\\
  0.47  &   0.08460(7)[+5]  & 1.17991(1)[-9] & 0.28224(12)[-24]\\
  0.48  &   0.11591(7)[-25] & 1.27155(5)[-27]& 0.38405(15)[-20]\\
  0.50  &   0.16644(8)[-8]  & 1.42298(3)[-21]& 0.53885(16)[-31]\\
  0.52  &   0.20886(5)[-11] & 1.54594(2)[-17]& 0.65719(18)[-2]\\
  0.525 &   0.21859(6)[-2]  & 1.57336(3)[-10]& 0.68335(4)[-12] \\  
  0.55  &   0.26344(6)[-14] & 1.69440(4)[-24]& 0.79552(39)[-10]\\
\hline
\end{tabular}
\end{center}
\end{table}

\begin{figure}
\vskip1.0cm
\includegraphics[height=10.5cm]{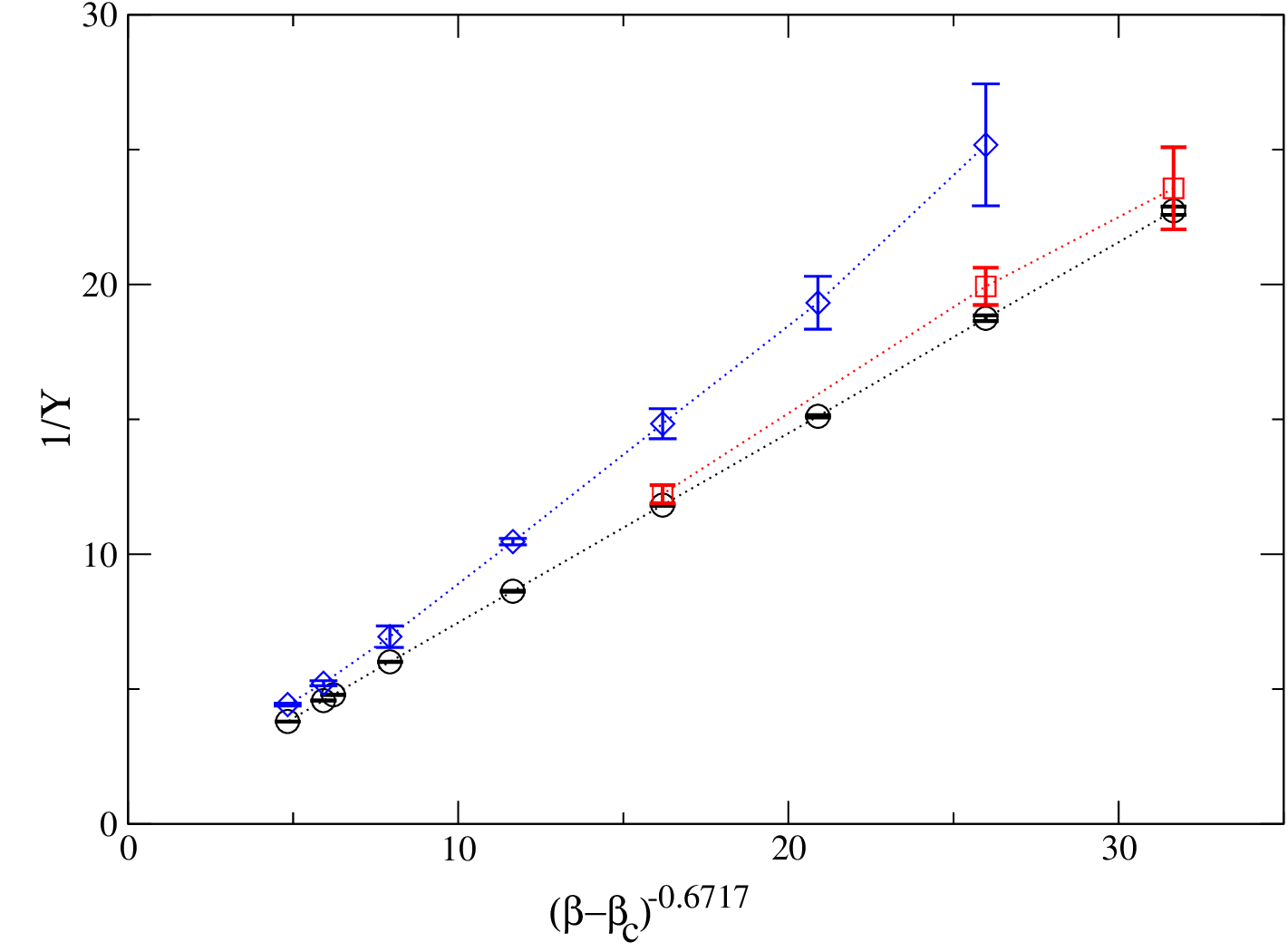}
\vskip0.3cm
\caption{
The inverse of the helicity modulus $\Upsilon$ is plotted
as a function of $(\beta_c-\beta)^{-\nu}$ using the numerical
value $\nu=0.6717$ for the exponent.  Our results are given by the black
circles; the  data
taken from ref. \cite{ToYo94} are represented by red squares
and the results of ref. 
\cite{Bielefeld2} are given by blue diamonds. The dotted lines should only 
guide the eye.
\label{helivgl}
}
\end{figure}

Our results for the helicity 
modulus for the XY model 
can be compared with those of refs. \cite{ToYo94,Bielefeld2}.
We have taken the values of $F$ from table 6 of ref. \cite{Bielefeld2}.
The relation with the helicity modulus, as defined here, is $\Upsilon = F^2$.
In the case of ref. \cite{ToYo94} we have taken the same quantity from their
table V.
In figure \ref{helivgl}  we have plotted the inverse of the helicity modulus 
as a function of $(\beta_c-\beta)^{-\nu}$ using the numerical 
value $\nu=0.6717$ for the exponent. 
The results of ref. \cite{ToYo94} are in reasonable agreement with ours. 
On the other hand, those of ref. \cite{Bielefeld2} clearly deviate.
In particular for small reduced 
temperatures, the value of the helicity modulus seems to be underestimated
in ref. \cite{Bielefeld2}. 
Both ref. \cite{ToYo94} and \cite{Bielefeld2} did not directly measure 
the helicity modulus using eq.~(\ref{helimeasure}) but instead extracted 
it from the $L$-dependence of the magnetisation. In the case of ref. 
\cite{Bielefeld2}, in addition, the simulations were done for a finite 
external field, involving the extrapolation to vanishing external
field. 

\section{Simulations in the high temperature phase}
In the high temperature phase, we expect that the observables converge
exponentially fast towards the thermodynamic limit. Therefore we have
used lattice sizes that are large enough  to ensure that the
systematic error due to the finite lattice size is by far smaller 
than the statistical error of the observables that we have measured. 
We have checked that this is fulfilled for  $L \gtrapprox 10 \xi_{2nd}$.
In fact for most of our the simulations we have chosen $L > 12 \xi_{2nd}$. 
Most of the simulations reported here were already performed in 
connection with refs. \cite{newXY,MHHe}. 
The magnetic 
susceptibility $\chi$ and $F$ were measured using so called improved 
estimators; see e.g. ref. \cite{uwolff89}.  
 Note that these improved estimators are, in contrast
to the naive ones, self-averaging.  Hence, for large $L/\xi$ their
statistical error is much smaller than that of the naive estimators.
For the largest values of $\beta$ we have performed $2 \times 10^5$ 
measurements. For each measurement we performed one Metropolis
sweep and $n_c$ single cluster updates. The number $n_c$ of 
single cluster updates was chosen such that the average size of a
cluster times $n_c$ is roughly equal to the number of sites of the lattice.
The simulations for the largest lattice size $L=350$ took about one month
on a 2 GHz Opteron CPU for each value of $\lambda$.
Our results for the high temperature phase of the $\phi^4$ model at 
$\lambda=2.1$ and $\lambda=2.2$ and the XY model 
are summarized in tables \ref{highphi4}, \ref{highl2.2} and \ref{highXY},
respectively.  
\begin{table}
\caption{\sl \label{highphi4}
Results for the high temperature phase of the 2-component $\phi^4$ model at
$\lambda=2.1$. $\beta$ is the inverse temperature, $\xi_{2nd}$ the second
moment correlation length, $\chi$ the magnetic susceptibility,
$E$ the energy and $C$ the specific heat. For the definition of these quantities
and detailed discussion see the text.
}
\begin{center}
\begin{tabular}{|l|l|l|l|l|}
\hline
  \multicolumn{1}{|c}{$\beta$}
& \multicolumn{1}{|c}{$\xi_{2nd}$}
& \multicolumn{1}{|c}{$2 \chi$}
& \multicolumn{1}{|c}{$E$} 
& \multicolumn{1}{|c|}{$C$} \\
\hline
0.40  &\phantom{0}1.07551(2) &\phantom{000}5.7056(2) &0.512689(10)&                                                                            \phantom{0}2.091(3) \\
0.41  &\phantom{0}1.15844(2) &\phantom{000}6.4486(2) &0.534164(10)&
                                                          \phantom{0}2.213(3) \\
0.42  &\phantom{0}1.25591(3) &\phantom{000}7.3867(3) &0.556926(12)&
                                                          \phantom{0}2.349(3) \\
0.43  &\phantom{0}1.37262(3) &\phantom{000}8.6013(4) &0.581128(13)&
                                                          \phantom{0}2.500(4) \\
0.44  &\phantom{0}1.51582(4) &\phantom{00}10.2280(5) &0.607060(12)&
                                                          \phantom{0}2.692(4) \\
0.45  &\phantom{0}1.69738(4) &\phantom{00}12.5045(5) &0.635073(7) &
                                                          \phantom{0}2.902(7) \\
0.455 &\phantom{0}1.80821(4) &\phantom{00}14.0117(5) &0.649978(6) &
                                                          \phantom{0}3.050(8) \\
0.46  &\phantom{0}1.93718(5) &\phantom{00}15.8770(8) &0.665592(8) &
                                                          \phantom{0}3.199(8) \\
0.465 &\phantom{0}2.08960(6) &\phantom{00}18.2377(8) &0.682020(8) &
                                                          \phantom{0}3.377(7) \\
0.47  &\phantom{0}2.27341(8) &\phantom{00}21.3070(13)&0.699359(9) &
                                                          \phantom{0}3.580(9) \\
0.475 &\phantom{0}2.50035(11)&\phantom{00}25.4333(18)&0.717754(9) &
                                                          \phantom{0}3.813(8) \\
0.48  &\phantom{0}2.79047(12)&\phantom{00}31.2425(22)&0.737430(9) &
                                                          \phantom{0}4.081(10)\\
0.485 &\phantom{0}3.17678(14)&\phantom{00}39.909(3)  &0.758610(9) &
                                                          \phantom{0}4.385(11)\\
0.49  &\phantom{0}3.72370(19)&\phantom{00}53.998(5)  &0.781682(7) &
                                                           \phantom{0}4.87(2) \\
0.493 &\phantom{0}4.1825(2)  &\phantom{00}67.451(6)  &0.796679(6) &
                                                           \phantom{0}5.09(3)  \\
0.495 &\phantom{0}4.5763(2)  &\phantom{00}80.177(6)  &0.807259(6) &
                                                           \phantom{0}5.42(2)  \\
0.50  &\phantom{0}6.1498(5)  &\phantom{0}141.899(17) &0.836367(6) &
                                                           \phantom{0}6.26(2)  \\
0.503 &\phantom{0}8.0424(8)  &\phantom{0}238.88(5)  &0.856373(7) &
                                                           \phantom{0}7.14(5)  \\
0.505 &          10.4822(16) &\phantom{0}400.38(11) &0.871351(8) &
                                                           \phantom{0}7.95(3)  \\
0.506 &          12.6264(16) &\phantom{0}575.74(14) &0.879580(6) &
                                                           \phantom{0}8.52(5)  \\
0.507 &          16.318(4)   &\phantom{0}950.7(4)   &0.888476(7) &
                                                           \phantom{0}9.43(8)  \\
0.5075&          19.498(6)   &          1347.1(8)   &0.893283(9) &
                                                           \phantom{0}9.83(7)  \\
0.508 &          24.845(8)  &          2164.6(1.4)  &0.898418(6) &10.81(8)\\
0.5083&          30.453(10)  &         3225.0(2.0)  &0.901727(4) &11.32(9)\\
\hline
\end{tabular}
\end{center}
\end{table}

\begin{table}
\caption{\sl \label{highl2.2}
Same quantities as in table \ref{highphi4} for $\lambda=2.2$.
}
\begin{center}
\begin{tabular}{|l|l|l|l|l|}
\hline
  \multicolumn{1}{|c}{$\beta$}
& \multicolumn{1}{|c}{$\xi_{2nd}$}
& \multicolumn{1}{|c}{$2 \chi$}
& \multicolumn{1}{|c}{$E$}
& \multicolumn{1}{|c|}{$C$} \\
\hline
 0.501   & \phantom{0}7.1723(5)& \phantom{0}191.372(23)& 0.849150(4)  &  \phantom{0}6.681(10) \\
 0.5035  & \phantom{0}9.5018(9) & \phantom{0}330.80(6)  & 0.866864(5)  &  \phantom{0}7.545(15) \\
 0.5055  & 13.6104(21)&  \phantom{0}667.11(20) & 0.882972(5)  &  \phantom{0}8.62(4)   \\ 
 0.5067  & 19.710(5)  & 1376.2(6)   & 0.894014(6)  &  \phantom{0}9.87(6)   \\
 0.50748 & 30.475(10) & 3231.4(2.0) & 0.902171(4)  & 11.41(9)   \\
\hline
\end{tabular}
\end{center}
\end{table}

\begin{table}
\caption{\sl \label{highXY}
Same quantities as in table \ref{highphi4} for the standard XY model.
}
\begin{center}
\begin{tabular}{|l|l|l|l|l|}
\hline
  \multicolumn{1}{|c}{$\beta$}
& \multicolumn{1}{|c}{$\xi_{2nd}$}
& \multicolumn{1}{|c}{$2 \chi$}
& \multicolumn{1}{|c}{$E$}
& \multicolumn{1}{|c|}{$C$} \\
\hline
0.4  &\phantom{0}1.87550(7) &\phantom{0}16.8465(10)&0.740797(15) & 3.139(4) \\
0.41 &\phantom{0}2.18009(8) &\phantom{0}22.0757(13)& 0.773412(11)& 3.388(5) \\
0.42 &\phantom{0}2.62528(12)&\phantom{0}31.019(2) & 0.809006(12) & 3.736(6) \\
0.425&\phantom{0}2.93916(12)&\phantom{0}38.242(2) & 0.828195(8)  & 3.957(6) \\
0.43 &\phantom{0}3.35803(15)&\phantom{0}49.064(4) & 0.848624(8)  & 4.231(7) \\  
0.435&\phantom{0}3.9514(2)  &\phantom{0}66.702(5) & 0.870561(8)  & 4.561(6) \\
0.44 &\phantom{0}4.8769(2)  &\phantom{0}99.564(6) & 0.894452(5)  & 5.014(8) \\
0.441&\phantom{0}5.1305(2)  &          109.704(8) & 0.899537(6)  & 5.139(7) \\
0.442&\phantom{0}5.4184(2)  &          121.801(8) & 0.904735(5)  & 5.242(8) \\
0.443&\phantom{0}5.7489(3)  &          136.464(12)& 0.910056(6)  & 5.390(7) \\
0.444&\phantom{0}6.1327(3)  &          154.524(14)& 0.915514(6)  & 5.542(7) \\
0.445&\phantom{0}6.5851(4)  &          177.23(2)  & 0.921133(6)  & 5.704(8) \\
0.446&\phantom{0}7.1286(5)  &          206.53(2)  & 0.926937(6)  & 5.874(10)\\
0.447&\phantom{0}7.7960(4)  &          245.53(2)  & 0.932911(4)  & 6.104(9) \\
0.448&\phantom{0}8.6402(4)  &          299.62(2)  & 0.939131(3)  & 6.342(9) \\
0.449&\phantom{0}9.7488(6)  &          378.69(4)  & 0.945605(4)  & 6.644(10)\\
0.45 &          11.2871(7)  &          503.40(5)  & 0.952402(3)  & 6.959(10)\\
\hline
\end{tabular}
\end{center}
\end{table}

\section{Universal amplitude ratios}
In this section, we first define the amplitude ratios
that we consider. Then we compute their values from the Monte Carlo
data discussed in the previous sections.  Finally we compare our results
with those obtained by using other theoretical methods and experimental 
results obtained for the $\lambda$-transition of $^4$He.
\subsection{Definition of the amplitude ratios}
Here we define the amplitude ratios that we have studied. For a more 
comprehensive list see e.g. page 18 of ref. \cite{PrHoAh91} or 
table 2 of ref. \cite{review}. The first amplitude
combination relates the correlation length in the high temperature phase with 
the helicity modulus which is defined in the low temperature phase:
\begin{equation}
 R_{\Upsilon} \equiv  f_{2nd,+} \Upsilon_0 \;\;,
\end{equation}
where the amplitudes are defined by
\begin{equation}
\label{xiU}
\xi_{2nd} = \simeq f_{2nd,+}  (-t)^{-\nu} \;\;, \;\; \;\;  \Upsilon  \simeq  \Upsilon_0 t^{\nu} \;\;,  \;\; \;\; (t>0)
\end{equation}
where $\nu$ is the critical exponent of the correlation length. In this section we use,
for computational convenience,
\begin{equation}
\label{reduced}
 t=\beta-\beta_c
\end{equation}
as definition  of the reduced temperature. It has been
shown in ref. \cite{FiBaJa73} that the helicity modulus behaves as an 
inverse correlation length. 
The exponential correlation length $\xi_{exp}$,
which describes the asymptotic decay of the correlation function, 
differs only slightly from $\xi_{2nd}$ which has been used here. Following
ref. \cite{ourXY}:
\begin{equation}
 \lim_{t\rightarrow 0} \frac{\xi_{exp}}{\xi_{2nd}} = 1.000204(3)  \;\;,\;\;\;\; (t<0) 
 \;\;.
\end{equation}

Next we consider
\begin{equation}
\label{RB}
R_B \equiv \frac{C_+}{f_{2nd,+}^3 B^2} \;\;,
\end{equation}
where the amplitudes of the magnetic susceptibility and the correlation length
in the high temperature phase are defined by eqs.~(\ref{chipower},\ref{xiU}),
respectively.  
The amplitude of the magnetisation in the low temperature phase is defined by
\begin{equation}
\label{magpower}
\Sigma/2  \simeq \left(B t^{\beta} \right)^2 \;\;,\;\;\;\; (t>0) \;\;.
\end{equation}

As the third combination of amplitudes we consider
\begin{equation}
\label{Rxip}
 R_{\xi}^+ \equiv (\alpha A^+)^{1/3} f_{2nd,+} \;,  
\end{equation}
where the amplitude of the specific heat $C_h$ is given by
\begin{equation}
\label{heatpower}
C_h  \simeq A_{\pm} |t|^{-\alpha} + b \;.
\end{equation}
The analytic background $b$ has to be taken into account, since the exponent
of the specific heat $\alpha$ is negative for the three-dimensional XY universality 
class.
The fourth combination of amplitudes that we consider is
\begin{equation}
\label{Rxim}
 R_{\xi}^- \equiv \frac{(\alpha A^-)^{1/3}}{\Upsilon_0} \;\;.
\end{equation}

In the case of $R_{\xi}^+$ and $R_{\xi}^-$ one has to be careful about the 
precise definition of the specific heat and the reduced temperature. We have checked
that, due to a cancellation, our definitions of $R_{\xi}^+$ and $R_{\xi}^-$ indeed 
coincide with those used in the literature.

\subsection{Numerical results}
\begin{figure}
\includegraphics[height=10.5cm]{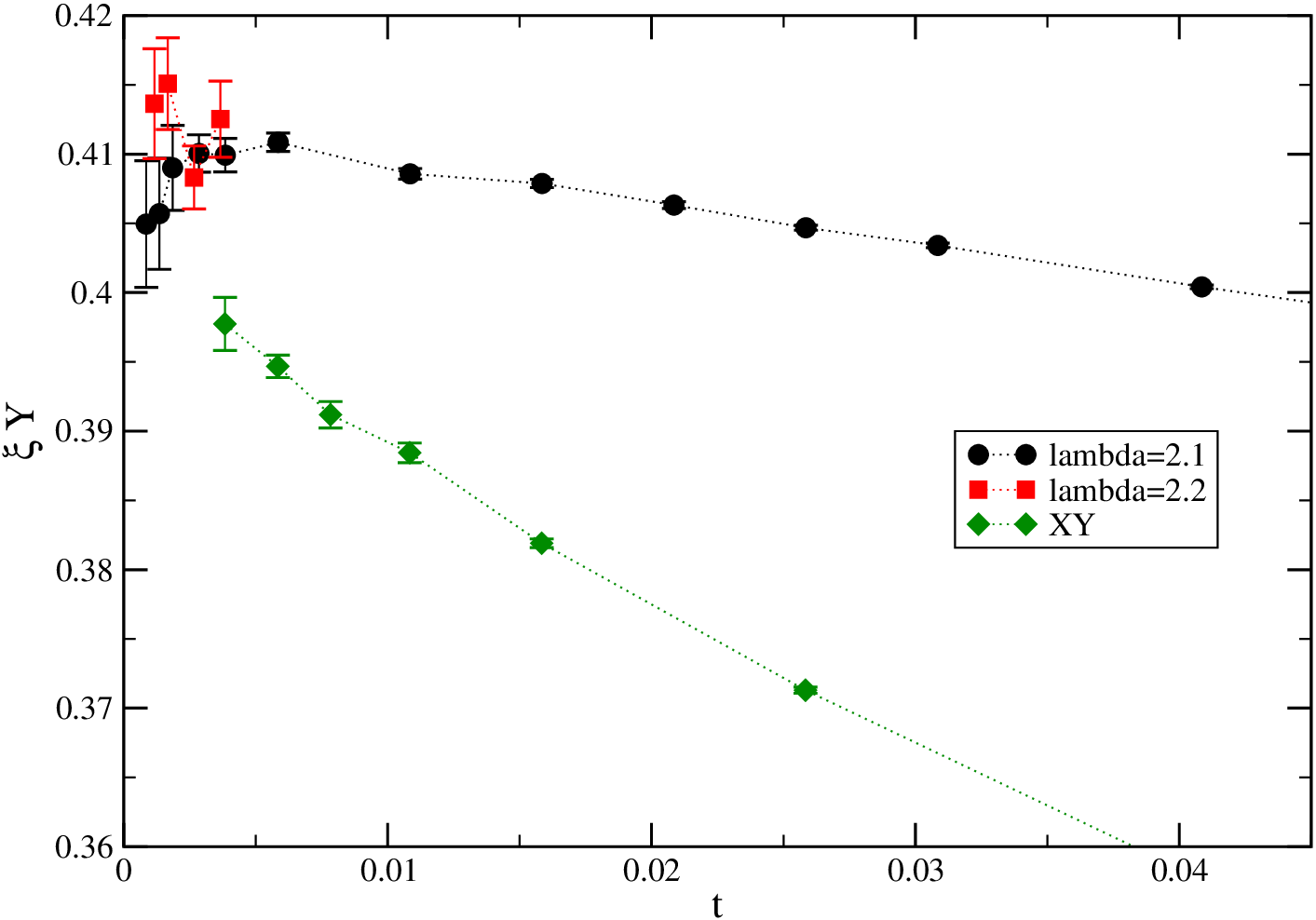}
\vskip0.3cm
\caption{
The product $\xi_{2nd}(-t) \Upsilon(t)$ for 
the $\phi^4$ model at $\lambda=2.1$, $\lambda=2.2$ and the XY model is  plotted
as a function of $t=\beta-\beta_c$, where 
we have taken $\beta_c=0.5091503$, $0.5083355$ and $0.4541655$ for 
the $\phi^4$ model at $\lambda=2.1$, $\lambda=2.2$ and the XY model, respectively.
The dotted lines should only guide the eye.
\label{xiheli}
}
\end{figure}

A straight forward way to obtain numerical estimates of amplitude ratios is to fit 
the numerical data for the various observables using ans\"atze  like
eqs.~(\ref{chipower},\ref{xiU},\ref{magpower},\ref{heatpower}).  Then the 
amplitude ratios are simply computed from the amplitudes that are obtained 
from these fits. Here instead we follow a strategy
that had already been employed in ref. \cite{isingratio} where universal 
amplitude ratios for the three-dimensional Ising universality class had been computed. 
In order to eliminate the dependence of the result on the critical
exponents, we consider ratios at a finite reduced temperature~(\ref{reduced}).
As a first example let us consider 
\begin{equation}
 R_{\Upsilon}=  \lim_{t \rightarrow 0}  \xi_{2nd}(-t) \Upsilon(t) \;\;,\;\;\;\;(t>0)\;\;.
\end{equation}
We have computed this product for the reduced temperatures $t=\beta_{low}-\beta_c$,  
where  $\beta_{low}$ are those values of $\beta > \beta_c$ for which  simulations
have been performed. The values for $\Upsilon$ are taken from tables \ref{lowphi4},
\ref{lowphi4l2.2} and \ref{lowXY}.
The tables \ref{highphi4}, \ref{highl2.2} and
\ref{highXY} contain no exact matches for $-t=\beta_c-\beta_{low}$. Therefore
we computed  $\xi_{2nd}$ at $-t$, i.e. at 
$\beta_{high}=2 \beta_c -\beta_{low}$, by interpolating  the 
results given in tables \ref{highphi4}, \ref{highl2.2} and 
\ref{highXY}. For this purpose, we took two values $\beta_{high,1}$ and 
$\beta_{high,2}$, where we have simulated, such that 
$\beta_{high,1} \le  \beta_{high} \le  \beta_{high,2}$ and 
$\beta_{high,2}-\beta_{high,1}$ is minimal. Then we interpolate linearly in
$f_{2nd}(\beta) = \xi_{2nd}(\beta) (\beta_c-\beta)^{0.6717}$ to get
$\xi_{2nd}$ at  $\beta_{high}$.
Our results for
$\xi_{2nd}(-t) \Upsilon(t)$ are shown in figure \ref{xiheli}. The statistical error
of the product is completely dominated by the error of $\Upsilon$.  Unfortunately,
the error rapidly increases as we approach the critical temperature. We have
checked that the error of $\xi_{2nd}(-t) \Upsilon(t)$ induced by the
uncertainty of $\beta_c$ is negligible.

Based on RG theory (see e.g. ref. \cite{review}), we expect that the product behaves as
\begin{equation}
  \xi_{2nd}(-t) \Upsilon(t) = R_{\Upsilon} \; + \; b_1 t^{\theta} + c_1 t +  
                              d_1 t^{\theta'} + b_2 t^{2 \theta} + ... \;\;.
\end{equation}
The leading non-analytic correction is characterized by the exponent 
$\theta =\nu \omega$.
As numerical values we take $\nu=0.6717(1)$ and $\omega=0.785(20)$ 
given in  ref. \cite{newXY}.  This value of $\omega$ is consistent with e.g. the result
of ref. \cite{GZ-98} $\omega=0.789(11)$ from the perturbative expansion in $d=3$ and
$\omega=0.802(18)$ from the $\epsilon$-expansion.
There is little information on $\theta'$ in the literature. We assume the value
$\theta' \approx 2 \theta$ as obtained in ref. \cite{RG}. Since
$2 \theta \approx 1$, it is difficult to disentangle the three terms 
$c_1 t$, $d_1 t^{\theta'}$ and $b_2 t^{2 \theta}$. Therefore we have used 
the ansatz
\begin{equation}
\label{ansatzxiheli}
  \xi_{2nd}(-t) \Upsilon(t) = R_{\Upsilon} \; + \; b t^{\theta} + c t   
\end{equation}
for our fits.
Since $\lambda=2.1$ and $2.2$ are close to $\lambda^*$ we expect that 
$b$ is small for these values of $\lambda$. 
We have fitted our results for the $\phi^4$ model at $\lambda=2.1$ 
and the XY model with the ansatz~(\ref{ansatzxiheli}). In both cases, all
available data points were included into the fit.  
In the case of $\lambda=2.2$ we have just averaged the available
results, since we have only results for small values of $t$.
The results of these fits are given in table \ref{fitxiheli}. 
In all cases  we get an $\chi/$d.o.f. close to 1. For $\lambda=2.1$, 
the coefficient $b$ of 
corrections $\propto t^{\theta}$ is within error-bars consistent with zero,
confirming that $\lambda=2.1$ is close to $\lambda^*$.  Within two standard deviations
the result for $R_{\Upsilon}$ is consistent among the three different models.
As our final result we quote
\begin{equation}
\label{ourxiratio}
R_{\Upsilon} = 0.411(2) \;\;,
\end{equation}
where the error-bar is chosen such that the  result for each of the models is 
covered.

\begin{table}
\caption{\sl \label{fitxiheli}
Fits of the product $\xi_{2nd}(-t) \Upsilon(t)$ using the 
ansatz~(\ref{ansatzxiheli}) in the case of the $\phi^4$ model at $\lambda=2.1$ and 
the XY model. In the case of the $\phi^4$ model at $\lambda=2.2$ we have just averaged
over all data that are available.
}
\begin{center}
\begin{tabular}{|c|l|c|c|c|}
\hline
\multicolumn{1}{|c}{model}
& \multicolumn{1}{|c}{$R_{\Upsilon}$}
& \multicolumn{1}{|c}{$b$}
& \multicolumn{1}{|c}{$c$}
& \multicolumn{1}{|c|}{$\chi/$d.o.f.} \\
\hline
$\lambda=2.1$  & 0.4118(8)   &\phantom{--}0.001(10) & --0.28(2) & 1.12 \\
$\lambda=2.2$  & 0.4092(11)  &     -               &    -      & 0.67 \\
XY             & 0.4097(18)  &--0.186(28)          & --0.43(8) & 0.22 \\
\hline
\end{tabular}
\end{center}
\end{table}

\begin{figure}
\includegraphics[height=10.5cm]{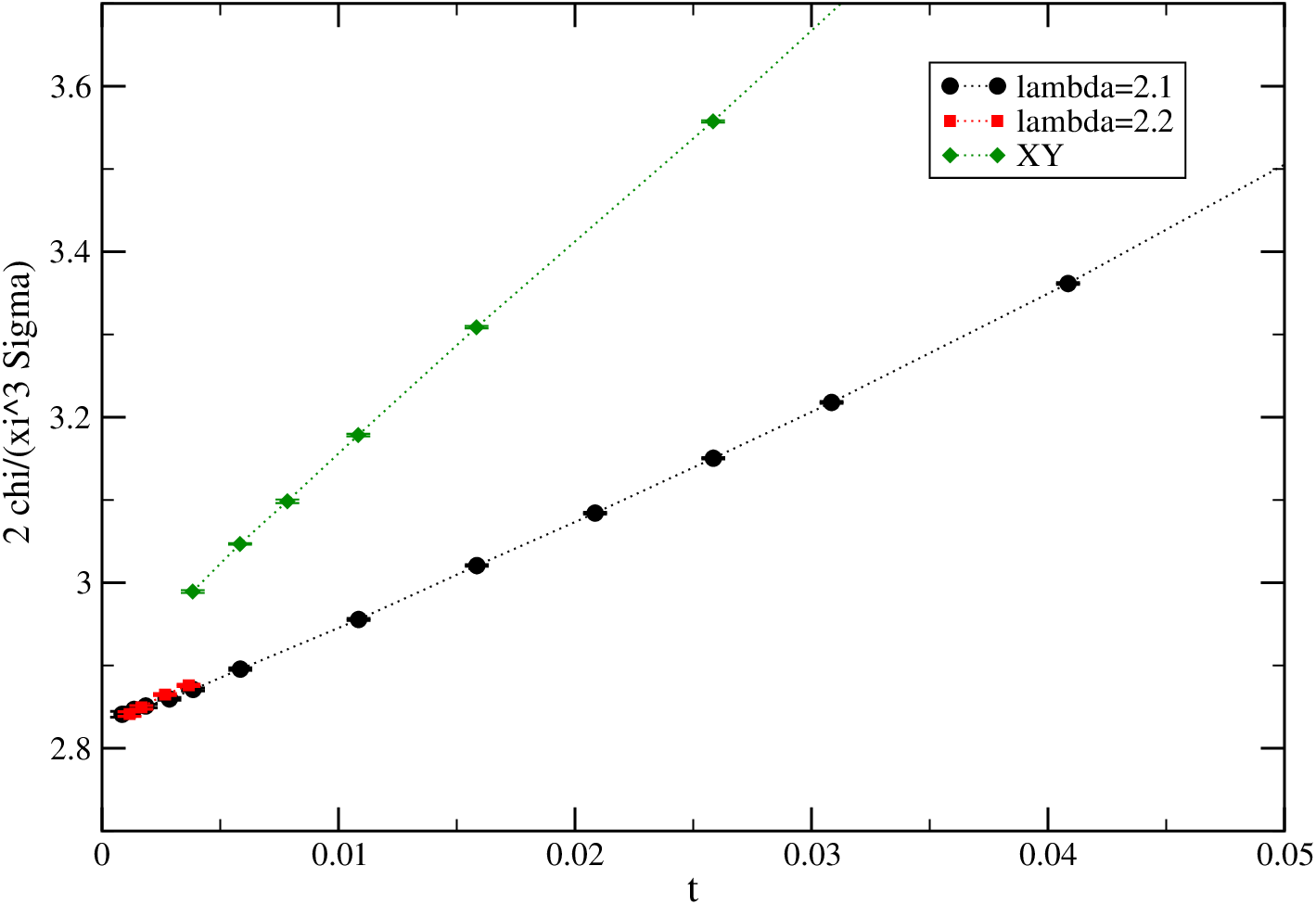}
\vskip0.3cm
\caption{
We plot the product $\frac{2 \chi(-t)}{\xi_{2nd}(-t)^3 \Sigma(t)}$ for
the $\phi^4$ model at $\lambda=2.1$, $\lambda=2.2$ and the XY model as
a function of $t=\beta-\beta_c$, where
we have taken $\beta_c=0.5091503$, $0.5083355$ and $0.4541655$ for
the $\phi^4$ model at $\lambda=2.1$, $\lambda=2.2$ and the XY model, respectively.
The dotted lines should only guide the eye.
\label{chixim_plot} 
}
\end{figure}

Next we have computed the amplitude ratio $R_B$. Also here we have computed 
the ratio of observables at finite reduced temperatures:
\begin{equation}
 R_B = \lim_{t \rightarrow  0} \frac{2 \chi(-t)}{\xi_{2nd}(-t)^3 \Sigma(t)} \;\;,
 \;\;\;\;(t>0) \;\;.
\end{equation}
To this end we have followed a similar approach as for $R_{\Upsilon}$.  We took 
the values $\Sigma(\beta_{low})$  given in tables \ref{lowphi4}, \ref{lowphi4l2.2} 
and  \ref{lowXY} for the $\phi^4$ model
at $\lambda=2.1$ and  $\lambda=2.2$ and the XY model, respectively. Then we computed 
$\xi_{2nd}(\beta_{high})$ and $\chi(\beta_{high})$ at 
$\beta_{high}= 2 \beta_c - \beta_{low}$ by interpolation of the results given  in 
tables \ref{highphi4}, \ref{highl2.2} and \ref{highXY}, for the $\phi^4$ model
at $\lambda=2.1$ and  $\lambda=2.2$ and the XY model, respectively. 
The interpolation is done in an analogous way as discussed above.
Our results are plotted in figure \ref{chixim_plot}.

In principle, a similar ansatz as for $R_{\Upsilon}$  could be used here. 
However it turns out that the analytic background of the magnetic susceptibility
has to be taken into account to get acceptable fits for a large range of 
$\beta$-values. Hence we have used the ansatz
\begin{equation}
\label{rchixim1}
 \frac{2 \chi(-t)}{\xi_{2nd}(-t)^3 \Sigma(t)} = R_B  + b t^{\theta} + c t  + d t^{\gamma}
 \;\;.
\end{equation}
The results of these fits are summarized in table \ref{fitR1}. 
\begin{table}
\caption{\sl \label{fitR1}
 Fits of the $\phi^4$ model at $\lambda=2.1$ and the XY model using the 
 ansatz~(\ref{rchixim1}).  In the case of the XY model all available data are used, 
 while in the case of the $\phi^4$ model all data up to $\beta=0.55$ are used;
 i.e. $\beta=0.58$ is skipped.  In these fits $\beta_c=0.5091503$ and 
 $\beta_c=0.4541655$  is used as input for the $\phi^4$ model at $\lambda=2.1$ and 
 the XY model, respectively.
}
\begin{center}
\begin{tabular}{|c|c|l|l|l|c|}
\hline
  \multicolumn{1}{|c}{model}
  & \multicolumn{1}{|c}{$R_B$}
  & \multicolumn{1}{|c}{$b$}
  & \multicolumn{1}{|c}{$c$}
  & \multicolumn{1}{|c}{$d$} 
  & \multicolumn{1}{|c|}{$\chi^2/$d.o.f.} \\
  \hline
$\lambda=2.1$ &  2.828(5)\phantom{0}  &0.14(20)&\phantom{0}7.2(2.1)&14.3(5)& 1.44 \\
   XY         & 2.801(15) & 3.2(5)   & --3.1(4.7) & 46.4(7.6)& 2.21\\
\hline
\end{tabular}
\end{center}
\end{table}
To check the uncertainty due to the error of $\beta_c$, we have repeated the 
fits for shifted values of $\beta_c$  (both for computing the combination
$\frac{2 \chi(-t)}{\xi_{2nd}(-t)^3 \Sigma(t)}$ as well as in the ansatz). 
We get $R_B=2.815(3)$, $2.797(16)$ using $\beta_c=0.5091509$ and 
$\beta_c=0.4541665$ for the $\phi^4$ model and the XY model, respectively.

We also performed fits for $\lambda=2.1$ and $2.2$ taking only analytic 
corrections into account:
\begin{equation}
\label{rchixim2}
 \frac{2 \chi(-t)}{\xi_{2nd}(-t)^3 \Sigma(t)} = R_B  +  c t \;\;.
 \end{equation}
The results of these fits are given in table \ref{fitR2}. 
\begin{table}
\caption{\sl \label{fitR2}
 Fits of the $\phi^4$ model at $\lambda=2.1$ and $\lambda=2.2$ 
using the ansatz~(\ref{rchixim2}).
 In the case of $\lambda=2.1$  we have taken into account all data up to
 $\beta=0.513$, while for $\lambda=2.2$ all available data have been used.
}
\begin{center}
\begin{tabular}{|c|c|l|c|}
\hline
 \multicolumn{1}{|c}{model}
 & \multicolumn{1}{|c}{$R_B$}
 & \multicolumn{1}{|c}{$c$}
 & \multicolumn{1}{|c|}{$\chi^2/$d.o.f.} \\
\hline
$\lambda=2.1$ & 2.8329(26) &\phantom{0}9.7(9)  & 0.24 \\ 
$\lambda=2.2$ & 2.8276(29) &  13.4(1.0)& 1.13 \\ 
\hline
\end{tabular}
\end{center}
\end{table}
Also here we have repeated the fits with shifted values of $\beta_c$. We get 
$R_B=2.83040(26)$  and $2.82494(29)$  using $\beta_c=0.5091509$ and 
$\beta_c=0.5083362$  for $\lambda=2.1$ and $\lambda=2.2$, respectively.
Based on the fits with the ansatz~(\ref{rchixim2}) we arrive at our final result:
\begin{equation}
R_B = 2.83(1) \;\;.
\end{equation}
The error which is quoted covers the statistical error, the error due to the 
uncertainty of $\beta_c$ and the error due to residual leading corrections.
The latter is estimated by the difference of the results for $\lambda=2.1$ and 
$\lambda=2.2$.

Finally we have computed $R_{\xi}^+$ and $R_{\xi}^-$.
To this end we have analysed our very accurate data for the energy density.
In the neighbourhood of the critical temperature, 
the energy density behaves as 
\begin{equation}
\label{enesingular}
 E(t)  = e_{ns} + c_{ns} t  + a_\pm t^{1-\alpha} + ...
\end{equation}
with $a_\pm (1-\alpha) = A_\pm$.  $e_{ns}$ and $c_{ns}$
are the non-singular contributions to the energy density and the specific heat 
at the transition temperature. Their numerical values had been accurately 
determined for the $\phi^4$ model at $\lambda=2.1$ and $2.2$ in ref. 
\cite{MHHe} using finite size scaling at the transition temperature.
These results are given by \cite{MHHe}:
\begin{equation}
\label{e0l2.1}
 e_{ns} = 0.913213(5) + 20 \times (\beta_c- 0.5091503)
          + 5 \times 10^{-7} \times (1/\alpha+1/0.0151)
\end{equation}
for $\lambda=2.1$ and
 \begin{equation}
 \label{e0l2.2}
 e_{ns} = 0.913585(5) + 20 \times (\beta_c-0.5083355)
     + 6 \times 10^{-7} \times (1/\alpha+1/0.0151)
\end{equation}
for $\lambda=2.2$.
The results for the non-singular part of the specific heat are \cite{MHHe}:
\begin{equation}
\label{c0l2.1}
 c_{ns} = 157.9(5) +  147000 \times (\beta_c-0.5091503)
 -2.1 \times (1/\alpha+1/0.0151)
 \end{equation}
 for $\lambda=2.1$ and
\begin{equation}
\label{c0l2.2}
c_{ns} = 155.6(4)  + 121000 \times (\beta_c-0.5083355)
-2.1 \times (1/\alpha+1/0.0151)
\end{equation}
for $\lambda=2.2$.  Corresponding results for the XY model are not available.

\begin{figure}
\includegraphics[height=10.5cm]{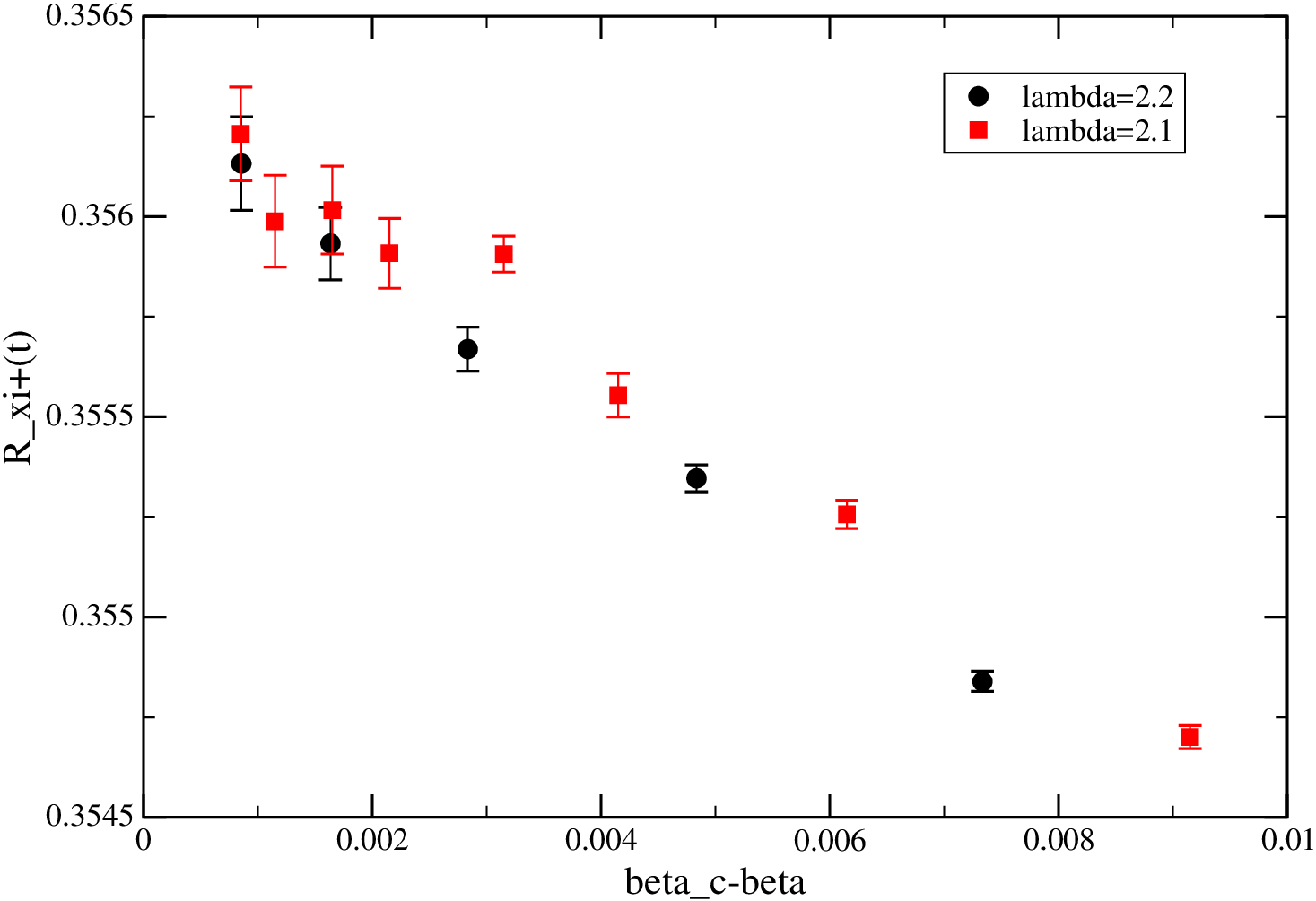}
\vskip0.3cm
\caption{
In this figure we show 
$R_{\xi}^+(t)=\xi_{2nd}(t)\left[-\alpha (1-\alpha)  e_s(t) (-t) \right]^{1/3}$ for
the $\phi^4$ model at $\lambda=2.1$ at $\lambda=2.2$.
This product is plotted as a function of $-t=\beta_c-\beta$, where
we have taken $\beta_c=0.5091503$ and $0.5083355$ for
the $\phi^4$ model at $\lambda=2.1$ and $\lambda=2.2$, respectively.
$e_{ns}$ and $c_{ns}$ are set to their central values. $\nu=0.6717$ is 
assumed.
\label{rxi_plot}
}
\end{figure}

Using these results we have computed the singular part of the energy density as
\begin{equation}
\label{esing}
e_s(\beta) = E(\beta)  - e_{ns} - c_{ns} (\beta-\beta_c) \;,
\end{equation}
where the numerical values for $E(\beta)$ are taken from tables \ref{highphi4} and
\ref{highl2.2} for the high temperature phase and 
 from tables \ref{lowphi4} and
\ref{lowphi4l2.2} for the low temperature phase.
First we have computed 
\begin{equation}
R_{\xi}^+  = \lim_{t \rightarrow 0}  
\xi_{2nd}(t)\left[-\alpha (1-\alpha)  e_s(t) (-t) \right]^{1/3} \;\;,\;\;\;\;  
(t<0) \;\;,
\end{equation}
for the high temperature phase.
Note that the combined critical exponent of the right hand side vanishes  due 
to the hyperscaling relation $d \nu = 2 - \alpha$. 
Our results for this combination are given in figure \ref{rxi_plot}.
In improved models, 
the corrections that are clearly visible, are to leading order due to terms 
$\propto t^2$ and $\propto t^{2-\alpha}$ missing in eq.~(\ref{enesingular}).
Our data do not allow to disentangle these two terms. Therefore we just linearly
extrapolated our results for 
$\xi_{2nd}(t)\left[-\alpha (1-\alpha)  e_s(t) (-t) \right]^{1/3}$
 to $t=0$.    We get from the extrapolation 
of the 5 largest values of $\beta$  the results $R_{\xi}^+=0.35616(11)$ 
and $R_{\xi}^+=0.35626(6)$
for $\lambda=2.1$ and $\lambda=2.2$, respectively. The error of $R_{\xi}^+$ is actually 
dominated by the error induced by the uncertainty of our input parameters, $e_{ns}$, 
$c_{ns}$, $\beta_c$ and $\alpha$.  In order to estimate this error, we have repeated
the whole procedure for shifted values of these input parameters. In 
particular, we have replaced $e_{ns}$ by $e_{ns}+\mbox{error}$ and similarly for the 
other input parameters. The errors obtained this way are very similar for the 
two values of $\lambda$. The largest contribution to the error originates from 
the uncertainty in $c_{ns}$ followed by $\beta_c$, $\alpha$ and $e_{ns}$. 
Note that the relatively small error due to the uncertainty of $\alpha$ is due 
to a cancellation 
of the variation of the $\alpha$ that appears explicitly in the definition of 
$R_{\xi}^+$ and that due to the dependence of $c_{ns}$~(\ref{c0l2.1},\ref{c0l2.2})
on $\alpha$.
Adding up all these errors we arrive at our final estimate
\begin{equation}
 R_{\xi}^+  = 0.3562(10)  \;\; .
\end{equation}

\begin{figure}
\includegraphics[height=10.5cm]{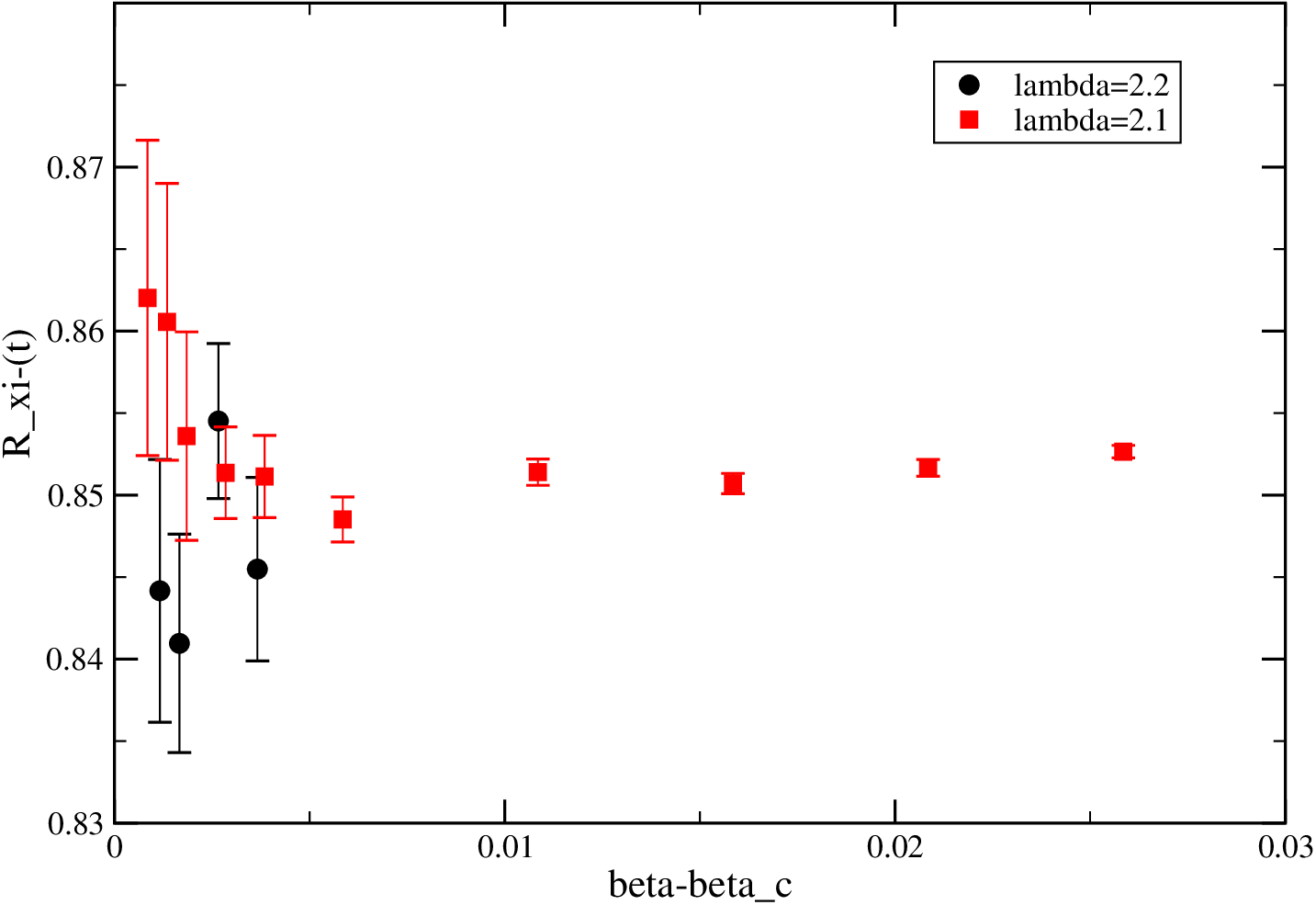}
\vskip0.3cm
\caption{
In this figure we give  
$R_{\xi}^-(t)=\Upsilon(t)^{-1} 
\left[-\alpha (1-\alpha)  e_s(t) \; (-t) \right]^{1/3}$ for
the $\phi^4$ model at $\lambda=2.1$ at $\lambda=2.2$.
It is plotted as a function of $t=\beta-\beta_c$, where
we have taken $\beta_c=0.5091503$ and $0.5083355$ for
the $\phi^4$ model at $\lambda=2.1$ and $\lambda=2.2$, respectively.
$e_{ns}$ and $c_{ns}$ are set to their central values. $\nu=0.6717$ is
assumed.
\label{rxim_plot}
}
\end{figure}

Finally we computed
\begin{equation}
R_{\xi}^-  = \lim_{t \rightarrow 0}
\frac{1}{\Upsilon(t)} \left[-\alpha (1-\alpha)  e_s(t) \; (-t) \right]^{1/3} 
\;\;,\;\;\;\; (t>0) \;\;.
\end{equation}
Note that in the low temperature phase $e_s(t)$ is negative, hence the 
product contained in $[]$ is positive.
Our numerical results for the $\phi^4$ model at $\lambda=2.1$ and $2.2$ are shown 
in figure \ref{rxim_plot}. Unfortunately the statistical error blows up for small
reduced temperatures. This is mainly due to the fact that the relative error of the 
helicity modulus $\Upsilon$ rapidly increases as the critical temperature is 
approached. In the case of $\lambda=2.1$, the value of $R_{\xi}^-$ stays essentially 
constant over the range $0.01  < \beta-\beta_c < 0.03$.  Therefore we regard  the
apparent increase of the value at small $\beta-\beta_c$ as  a statistical accident.
This is confirmed by the fact that for $\lambda=2.2$ no such trend can be seen.
Motivated by this consideration, 
we take our final result for  $R_{\xi}^-$ as the average over all data with 
$\beta-\beta_c<0.005$.  We get $R_{\xi}^-=0.852(2)$ and $0.848(3)$ for $\lambda=2.1$ 
and $\lambda=2.2$, respectively. As our final result we take the average over the 
two values of $\lambda$:
\begin{equation}
 R_{\xi}^-=0.850(5) \;\;.
\end{equation}
The error-bar is chosen such that the results of both values of $\lambda$, including 
their individual error-bars, are covered. We have also checked the possible error 
due to the uncertainty of the input parameters $\alpha$, $\beta_c$, $e_{ns}$ and 
$c_{ns}$.  Here, in contrast to $R_{\xi}^+$, these errors can be ignored. Actually 
the result for $R_{\xi}^-$ is virtually independent on $\alpha$. Using the experimental
value $\alpha=-0.0127$ \cite{LSNCI-96,lipa2003}
 instead of our theoretical one, the result
for $R_{\xi}^+$ changes only in the fourth digit.  This is due to the fact that
the variation of 
$\alpha$ that appears explicitly in the definition of $R_{\xi}^+$ essentially
cancels with that of $c_{ns}$~(\ref{c0l2.1},\ref{c0l2.2}).

\subsection{Comparison with other theoretical and experimental results}
In table \ref{compare} we have summarized results from the literature that
where obtained by different theoretical methods and from 
experiments on the $\lambda$-transition of $^4$ He. Essentially we followed
table 22 of ref. \cite{review}.

For $R_{\xi}^+$ most theoretical results are in reasonable agreement among each 
other. In the case of the field theoretic expansion in three
dimensions \cite{BeGo80,BaBe85} there is a discrepancy with our result that is 
somewhat larger than the combined error. There is no experimental result
for this amplitude ratio, since there is no direct experimental access to 
the correlation length of $^4$He in the high temperature phase. The authors
of ref. \cite{Bielefeld2} quote their Monte Carlo result as a function of
$\alpha$:
$R_{\xi}^+=0.3382(14) -0.717(96) \alpha + 0.87(1.13) \alpha^2$. We have 
inserted $\alpha=-0.0151$ to get the value quoted in table \ref{compare}.
It differs by about 3 times the combined error from our result.

In the case of $R_{\xi}^-$ our result is in perfect agreement with the 
experimental number given in table IV of ref. \cite{SiAh84}. It is interesting to 
note that also the experimental value of $R_{\xi}^-$ shows very little dependence
on the value of $\alpha$ that is assumed in the analysis; the authors of ref. 
\cite{SiAh84} arrive at $R_{\xi}^-=0.86$ and $R_{\xi}^-=0.84$ using 
$\alpha=-0.007$ and $\alpha=-0.025$, respectively, instead of their preferred value
$\alpha=-0.016$. They do not explicitly quote an error for their estimate 
of $R_{\xi}^-$.  In section III.B. they write however that $(R_{\xi}^-)^3$ is constant 
within $5\%$ over the entire range of pressures from 0 to 30 bars. In the review
\cite{PrHoAh91} the result of ref. \cite{SiAh84} is quoted as $R_{\xi}^-=0.85 \pm 0.02$.
Concerning theory, there is a discrepancy by about
twice the combined error with the field theoretic result  
\cite{BuStDo97,StMoDo00}. The error of the result obtained by the 
$\epsilon$-expansion \cite{Bervillier,Hohenberg,PrHoAh91} is quite large.
 The authors
of ref. \cite{Bielefeld2} quote their Monte Carlo result as
$R_{\xi}^-= 1.1580 -0.696 \alpha + 0.97 \alpha^2 \pm 0.036$. The  huge
difference compared with our result is likely due to the difference in the 
estimates of $\Upsilon$ as discussed in section \ref{lowtemp}.

\begin{table}
\caption{\sl \label{compare} We summarize results obtained by using different
theoretical methods and experiments on the $\lambda$-transition of $^4$He. 
MC denotes Monte Carlo simulations of lattice models, HT the high 
temperature series expansion of lattice models, IHT-PR the high temperature
series expansion of improved lattice models combined with the parametric 
representation of the equation of state, d=3 exp the field theoretic expansion
in three dimensions fixed and $\epsilon$ exp the $\epsilon$-expansion. 
For convenience, we have added our  present results in the second column 
of the table. A detailed discussion is given in the text.
}
{\footnotesize
\begin{center}
\begin{tabular}{|c|l|l|l|l|l|}
\hline
 \multicolumn{1}{|c}{}
 & \multicolumn{1}{|c}{MC}
 & \multicolumn{1}{|c}{HT,IHT-PR}
 & \multicolumn{1}{|c}{d=3 exp}
 & \multicolumn{1}{|c}{$\epsilon$ exp}
 & \multicolumn{1}{|c|}{Experiment} \\
\hline
$R_{\xi}^+$   &0.3562(10)&0.355(3) \cite{ourXY}& 0.3606(20) \cite{BaBe85,BeGo80} & 0.36 \cite{Bervillier} & \\
  & 0.349(5) \cite{Bielefeld2}    &0.361(4) \cite{BuCo99}&   &   & \\
$R_{\xi}^-$   &0.850(5)  & & 0.815(10) \cite{BuStDo97,StMoDo00}& 
1.0(2) \cite{Bervillier,Hohenberg,PrHoAh91}& 0.85(2) \cite{SiAh84} \\
 &1.180(36)  \cite{Bielefeld2} &  &   &   & \\
$R_c^+$   &0.128(2)  &0.127(6) \cite{ourXY}&0.123(3) \cite{StLaDo99}&
0.106 \cite{AbHi77,AbMa78,AhHo76} & \\
 & 0.128(4)  \cite{Bielefeld2} &  &   &   & \\
$R_{\Upsilon}$&0.411(2)  & &  & 0.27  \cite{Hohenberg}  & 0.39 \cite{GrAh73,Ah73,Hohenberg}\\
  &0.293(9) \cite{Bielefeld2} & &  &  0.33 \cite{Bervillier,OkabeIdeura}&0.41 \cite{GrAh73,Ah73,Bervillier}\\
\hline
\end{tabular}
\end{center}
}
\end{table}

In the literature our universal amplitude ratio $R_B$ is not discussed. Instead one 
finds  
\begin{equation}
R_c^+ = \frac{\alpha A_+ C_+}{B^2}  \;\;.\;
\end{equation}
It can be expressed  as
\begin{equation}
R_c^+ = (R_{\xi}^+)^3 R_B  = 0.128(2) \;\;.
\end{equation}
where the error is dominated by the error of $R_{\xi}^+$.   Our present
result is in good agreement with that obtained in ref. \cite{ourXY} using
the high temperature expansion of improved models in combination 
with the parametric representation of the equation of state.
It also agrees with  ref. \cite{Bielefeld2}, where a combination  of 
Monte Carlo simulations and the parametric representation of the equation 
of state was used. In ref.  \cite{Bielefeld2} $R_c^+$ is given as a function
of $\alpha$. The value in our table is at $\alpha=-0.0151$.
Again there 
is a discrepancy with the field theoretic result \cite{StLaDo99} that is 
slightly larger than the combined error. The $\epsilon$-expansion for 
$R_c^+$ has been computed up to O($\epsilon^2$)  \cite{AbHi77,AbMa78,AhHo76}.
The number given in the table is obtained by simply setting $\epsilon=1$.
There is no experimental result for $R_c^+$, since there is no experimental
access in $^4$He to the analogue of the magnetisation in spin models. 

In the case of $R_{\Upsilon}$ we see the largest differences among the 
results obtained by using theoretical methods.
The Monte Carlo result, quoted in eq.~(97) of ref. \cite{Bielefeld2}, differs
by more then 10 times the combined error from ours. This huge
difference can be traced back to the discrepancy in  $\Upsilon$ at
given values of $\beta$ as discussed in section \ref{lowtemp}.
Hohenberg et al. \cite{Hohenberg} have computed $R_{\Upsilon}$ by using
the $\epsilon$-expansion to O($\epsilon$).
Bervillier \cite{Bervillier} extended the calculation up to
O($\epsilon^2$). 
Okabe and Ideura \cite{OkabeIdeura} corrected the calculation of Bervillier,
which does however not change the numerical value,
and computed the $1/N$-expansion resulting in $R_{\Upsilon}= 0.14$ for
$N=2$.

There are no direct experimental results for the correlation length in the 
high temperature phase of $^4$He. Instead one might use the data for the 
specific heat in combination with the theoretical result for $R_{\xi}^+$ to 
arrive at the amplitude of the correlation length. This way, 
using $R_{\xi}^+=0.36$ and the experimental results of refs. \cite{GrAh73,Ah73},
Hohenberg et al. \cite{Hohenberg} arrive at $R_{\Upsilon}=0.39$. 
Bervillier noted, 
see section III.A. of \cite{Bervillier},  that there is an error 
in the experimental value of the amplitude of the transversal correlation length
used by Hohenberg et al. \cite{Hohenberg}. He arrives at the corrected value 
$R_{\Upsilon}=0.41$. It would certainly be worth while to redo this calculation
using most recent experimental data; 
e.g. those of ref. \cite{LSNCI-96,lipa2003} for the specific heat and 
our estimate of $R_{\xi}^+$.

\section{Summary and Conclusions}
In this paper we have computed universal amplitude ratios for the three 
dimensional XY universality class. These results are based on Monte Carlo
simulations of the three dimensional XY model and the $\phi^4$-model at 
$\lambda=2.1$ and $\lambda=2.2$. Note that these values of $\lambda$ are 
close to $\lambda^*=2.15(5)$ \cite{newXY}, where leading corrections to scaling 
vanish. We performed simulations in the low and the high temperature phase 
of these models. Extracting results for the thermodynamic limit, one has to 
take into account the effect of the Goldstone mode in the low temperature phase.
For a discussion see section \ref{goldstone}.

Our results for universal amplitude ratios for the three-dimensional 
XY universality class
are throughout more precise than previous theoretical results. They
are obtained in a rather direct way, making hidden systematic errors 
unlikely. When available, our results agree nicely with experimental 
ones obtained for the $\lambda$-transition of $^4$He, giving further 
confirmation to the fact that
this transition shares the three-dimensional XY universality class.

The numerical results obtained here for the correlation length $\xi_{2nd}$, 
the helicity modulus $\Upsilon$, the energy density $E$ and the specific 
heat set the stage also for the study of the specific heat 
or the Casimir force in confined geometries using improved models. 

\section{Acknowledgements}
I like to thank Andrea Pelissetto and Ettore Vicari for remarks on the
literature. The simulations were mainly carried out at the computer centre
of the Dipartimento di Fisica dell'Universit\`a di Pisa.
This work was partially supported by the DFG under grant No JA 483/23-1.

\end{document}